\documentclass[%
 aip,
 amsmath,amssymb,
preprint,%
 reprint,%
]{revtex4}

\usepackage{graphicx}
\usepackage{dcolumn}
\usepackage{bm}

\usepackage[utf8]{inputenc}
\usepackage[T1]{fontenc}
\usepackage{mathptmx}

\makeatletter
\def\@email#1#2{%
	\endgroup
	\patchcmd{\titleblock@produce}
	{\frontmatter@RRAPformat}
	{\frontmatter@RRAPformat{\produce@RRAP{*#1\href{mailto:#2}{#2}}}\frontmatter@RRAPformat}
	{}{}
}%
\makeatother
\begin{document}


\title{Quantum dot photoluminescence as a versatile probe\\ to visualize the interaction between plasma and nanoparticles on a surface}

\author{Z. Marvi}

\author{T. Donders}%

\author{M. Hasani}%

\author{G. Klaassen}%

\author{J. Beckers}

\affiliation{ 
	Department of Applied Physics, Eindhoven University of Technology, 5600 MB Eindhoven, The Netherlands\\
	Email: z.marvi@tue.nl
}%



\begin{abstract}
We experimentally demonstrate that the interaction between plasma and nanometer-sized semiconductor quantum dots (QDs) is directly connected to a change in their photoluminescence (PL) spectrum. This is done by taking in-situ, high resolution, and temporally-resolved spectra of the light emitted by laser-excited QDs on an electrically floating sample exposed to a low pressure argon plasma. Our results show a fast redshift of the PL emission peak indicating the quantum-confined Stark effect due to direct plasma-charging of these nanostructures and the substrate surface, while other plasma-induced (thermal and ion) effects on longer time scales could clearly be distinguished from these charging effects. The presented results and method open up novel pathways to direct visualization and understanding of fundamental plasma-particle interactions on nanometer length scales.


\end{abstract}

\maketitle

\section{introduction}
In the field of complex plasmas – i.e. ionized gases containing nanometer- to micrometer-sized particles – the charge of plasma-immersed particles is the parameter of most interest. This is because obtaining elementary knowledge regarding plasma-particle charging is the main key to understanding many fundamental processes such as those in the planetary rings of Saturn \cite{Goertz1988_Electrostatic}, in strongly coupled collective particle-particle interactions e.g. in plasma crystals \cite{PhysRevLett.73.652, Thoma2005_PlasmaCrystal} in earth laboratories \cite{hyde2013helical} and in the International Space Station \cite{pustylnik2016plasmakristall,wang2016dust,rosenfeld2017dust}, and in plasma-particle interactions in the form of dust density waves (DDWs) \cite{Tadsen2015_Self}.

From an application point of view, particle charging plays an essential role in in-situ plasma-assisted synthesis of nanometer- to micrometer-sized structures \cite{Wetering2013_void,Greiner2018_Diagnostics,Hundt2011_Real, Santos2020_Influence}, and plasma-particle interaction in nuclear fusion technology \cite{rubel2001dust,winter2004dust}. More recently, with the ever-shrinking length scales in nanomanufacturing, 
using the surface-charge-driven interaction between contaminating nanoparticles and photon-induced plasmas may become the only route to effective nanoparticle contamination control \cite{van2021plasma}. 

 Although it has drawn the attention of the entire research community, thus far the fundamental interaction between plasma and particles on the nm size scale has been largely unexplored. This is mostly because of the non-existence of experimental data needed to verify the few modeling efforts available in literature \cite{Cui1994_Fluctuations,matsoukas1996stochastic,mamunuru2017existence,santos2019electrostatic}. The reason for this lack of experimental data for particles of nm length scales is that, currently, all experimental methods to obtain information about the particle charge - such as the particle resonance method \cite{Melzer1994_Experimental,Carstensen2011_Mass,Jung2016_Resonance,carstensen2012charging}, force-balancing \cite{beckers2011microparticles}, particle interactions with electric fields \cite{Minderhout2019_Charge}, and Mie ellipsometry and dust density waves analysis \cite{Tadsen2017_Amplitude,Melzer2020_DDW}- are either directly or indirectly based on light scattering techniques which face a substantial loss of signal-to-noise in the nm regime due to the sixth-power-dependence of the signal intensity $I$ on the particle size $r$ ($I \propto r^6$).

\begin{figure}
	
		\includegraphics[width=1\textwidth]{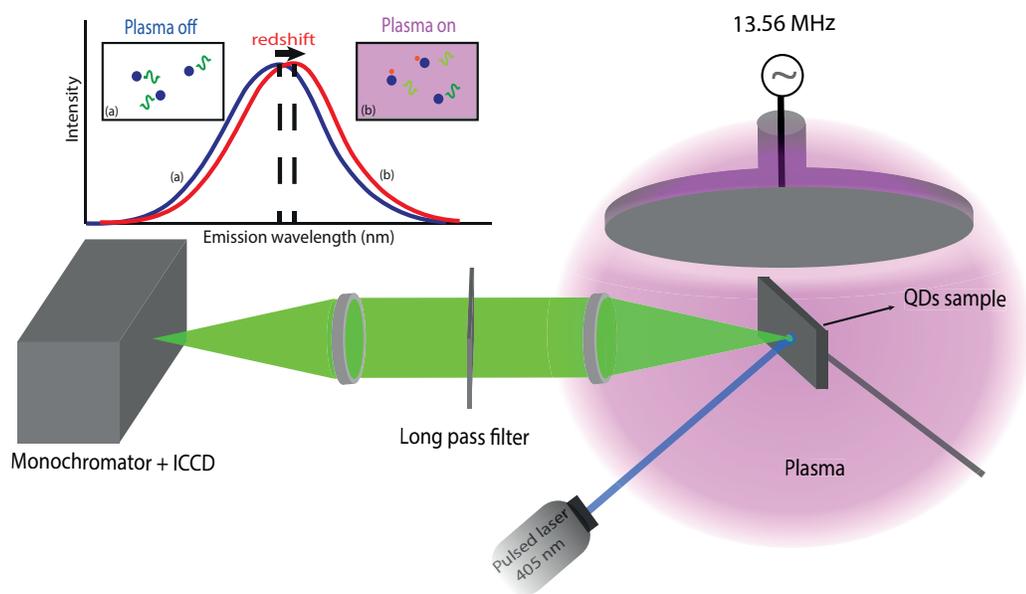} 
	\caption{(Color online) Schematic of the setup used to monitor the laser-induced photoluminescence of quantum dots deposited onto an electrically floating sample and the impact of plasma exposure on it.} \label{f3}
	
\end{figure}

In this letter, we demonstrate that the interaction between plasma and nm-sized semiconductor quantum dots (QDs) is directly connected to and can be visualized by a change in their photoluminescence emission spectrum. This is done by bringing an electrically floated QD-deposited sample into contact with a low pressure argon plasma and monitoring the photoluminescence (PL) spectrum of the QDs in a time-resolved fashion before, during, and after plasma exposure. The presented results and method open up new pathways to direct visualization and understanding of fundamental plasma-particle and plasma-surface interactions on nanometer length scales, and will allow researchers to obtain novel insights into all the aforementioned research fields and connected applications.  

\section{Methodology}
QDs behave as zero-dimensional
quantum wells for charge carriers. When these nanostructures are optically excited, an electron-hole pair is generated and confined in the quantum dot. This electron-hole pair relaxes back to the bound states in the quantum well, and, finally, recombines. During the latter process, a photon with a well-defined energy is emitted. When considering an ensemble of quantum dots this results in the emission of spectrally narrow-banded light, i.e. a photoluminescence peak.  
Essential in the method presented here is that the photoluminescence wavelength of the used QDs shifts due to the presence of charges and electric fields locally near their surface; a mechanism known as the quantum-confined Stark effect (QCSE) \cite{seufert2003single,park2012single}.

In our experiments, we used commercially available (from NanoOpticalMaterials) water-soluble CdSe/ZnS core-shell QDs with an average PL emission peak of $540\pm10$ nm and a full-width-at-half-maximum of 34 nm. Their core radius and the shell thickness were 2.1 nm and 0.9 nm, respectively. The QD samples were prepared by diluting colloidal QDs in distilled water to a concentration of 2 mg/ml, drop-casting 10 $\mu$l of that dilution on a silicon substrate and letting the liquid evaporate after which the QDs remained on the substrate. The QD sample was clamped on a stainless-steel substrate holder which was electrically insulated from the grounded walls. The QD sample, therefore, had a floating potential with respect to the plasma bulk. The photoluminescence emission upon laser excitation before, during, and after plasma exposure was recorded spectrally and temporally resolved with respective resolutions of 10 pm and 1.5 s. Figure~\ref{f3} shows a schematic of the experimental configuration developed to achieve this. 
  
The plasma discharge used for the exposure was driven by an RF generator set at a frequency of 13.56 MHz and an input RF power of 50 W. The axisymmetric circular RF electrode (12 cm in diameter) was mounted on top of the vacuum chamber. The chamber walls acted as a grounded counter electrode, which resulted in an asymmetric discharge. The vessel was equipped with a vacuum system, consisting of a prepump and a turbo-molecular pump, to achieve a base pressure of $10^{-4}$ Pa. Argon gas was fed into the chamber with a flow rate of $1$ sccm and the total gas pressure was kept constant for all measurements at $4$ Pa.   
  
 The QDs were excited in-situ using pulsed laser light with a wavelength of 405 nm, after which the photoluminescence emission of the QDs on the substrate was focused on the 250 micrometer entrance slit of a monochromator (Acton Research SpectraPro275). The spectrally separated light was then imaged by an ICCD camera (Andor iStar 334T) mounted on the exit slit of the monochromator. The output of the measurements is a  time series of spectra. By applying the excitation laser in a pulsed manner, the experiment allowed for direct background correction and averaging.

\section{results and discussion}
\begin{figure}
	
	\begin{center}
		\includegraphics[width=1\textwidth]{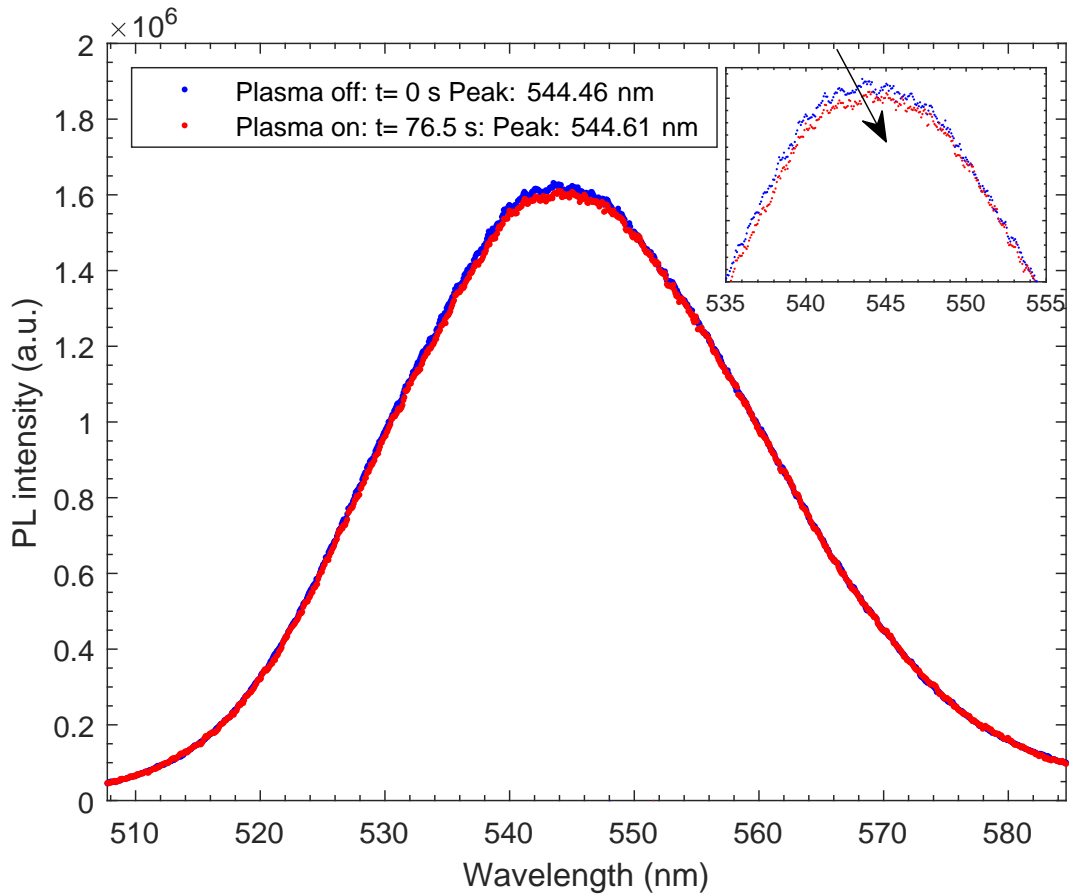} 

	\caption{(Color online) Photoluminescence spectrum emitted by the QDs before (blue) and during (red) plasma exposure. The plasma was operated at a pressure of 4 Pa and an input power of 50 W.} \label{f4}
		\end{center}
\end{figure}

Exemplary, figure~\ref{f4} shows two photoluminescence spectra from the same QD sample before (blue) and during (red) plasma exposure. A total redshift of 0.15 nm of the PL spectrum was observed after 76.5 s of plasma exposure. From spectral PL curves as shown in this figure, the center wavelength of the PL peak was determined, by fitting the background-corrected data with an exponentially modified Gaussian fit function, with an accuracy of 0.02 nm.

\begin{figure}
 	\begin{center}
 		 		\includegraphics[width=1\textwidth]{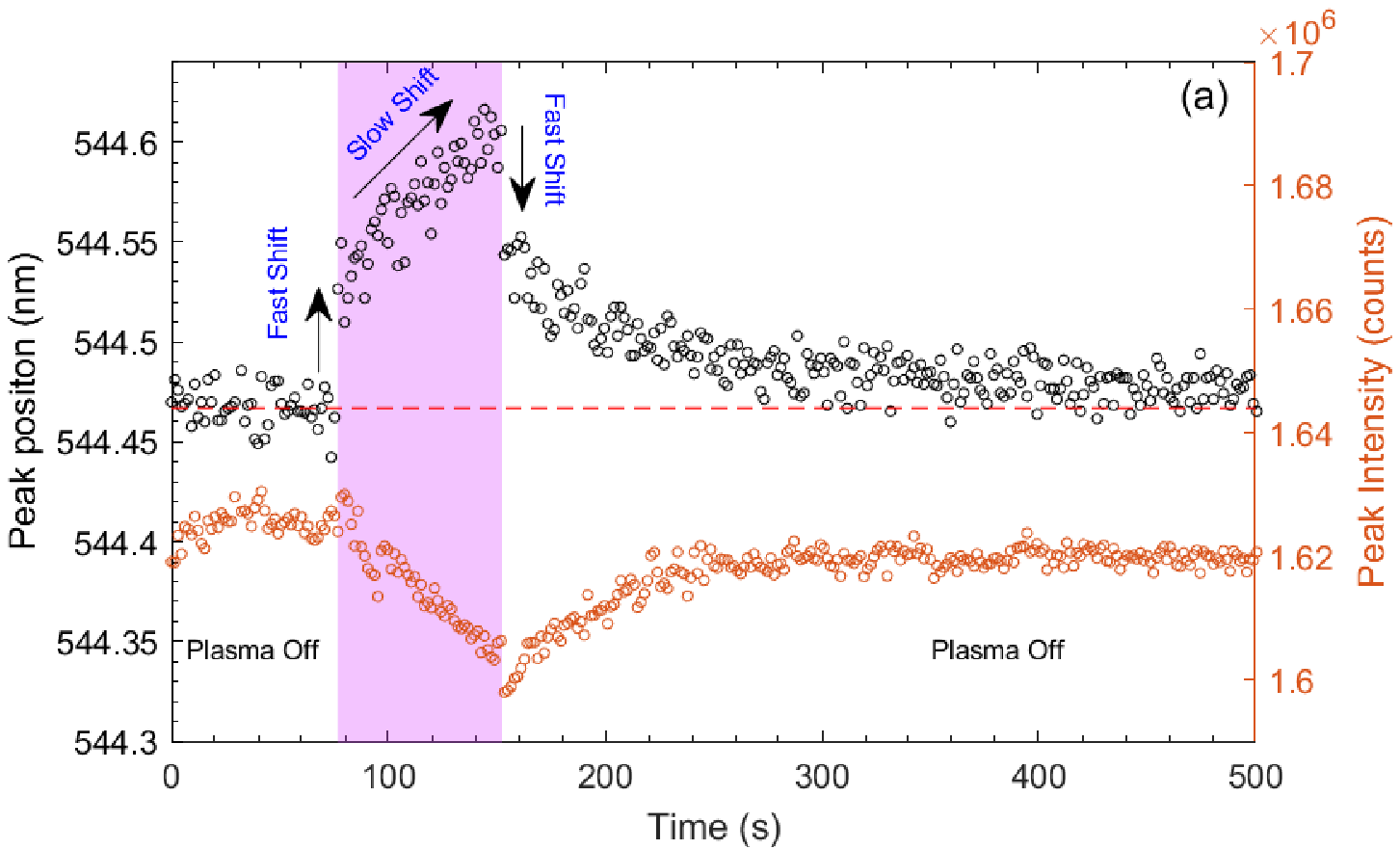} 

	\end{center}
 	\begin{center}
 		\includegraphics[width=1\textwidth]{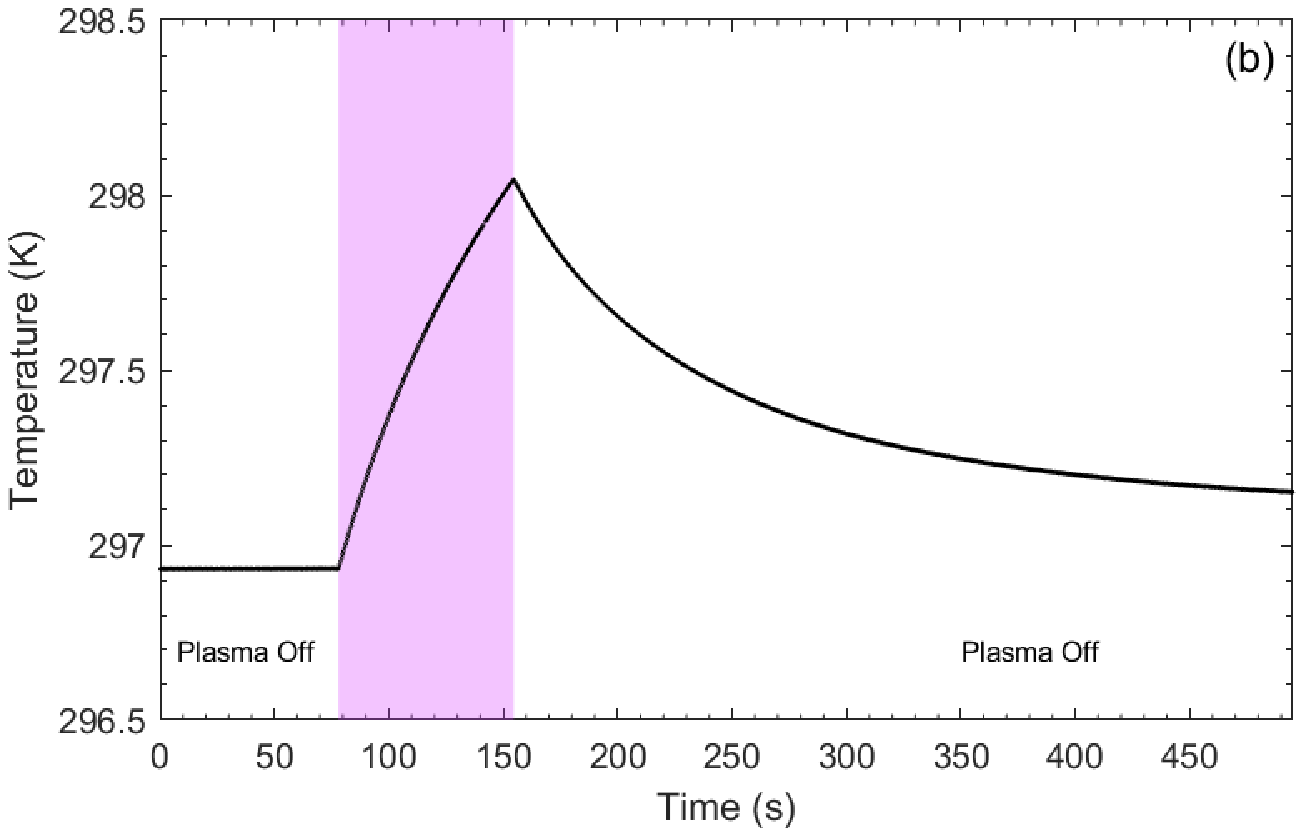} 
 	\end{center}
 \caption{(Color online) Time-traces of the center wavelength (left axis, black), and integrated intensity (right axis, red) of QDs PL peak (a), and the substrate temperature (b) during 76.5 s plasma exposure at input power of 50 W and pressure of 4 Pa. The shaded region indicates the plasma exposure.} \label{f5}
 \end{figure}

  Figure~\ref{f5}-a shows the center wavelength of the PL peak and the PL peak intensity as a function of time before, during, and after exposure of the QD sample to the plasma. As can be seen immediately from this figure, the PL peak's center wavelength responds to plasma impact on two distinctive time scales. On short timescales, an initial fast shift up to 0.04 nm within the first 1.5 s after the plasma was switched on is observed. On longer timescales, the center wavelength of the PL peak appears to shift slowly with time in the spectral domain over up to 0.11 nm during the remaining 75 s of plasma exposure. Moreover, the center wavelength of the PL peak appears to a fast shift back in the opposite direction of the initial redshift immediately after switching off the plasma.

 \begin{figure}
 
 	\begin{center} 

 		\includegraphics[width=1\textwidth]{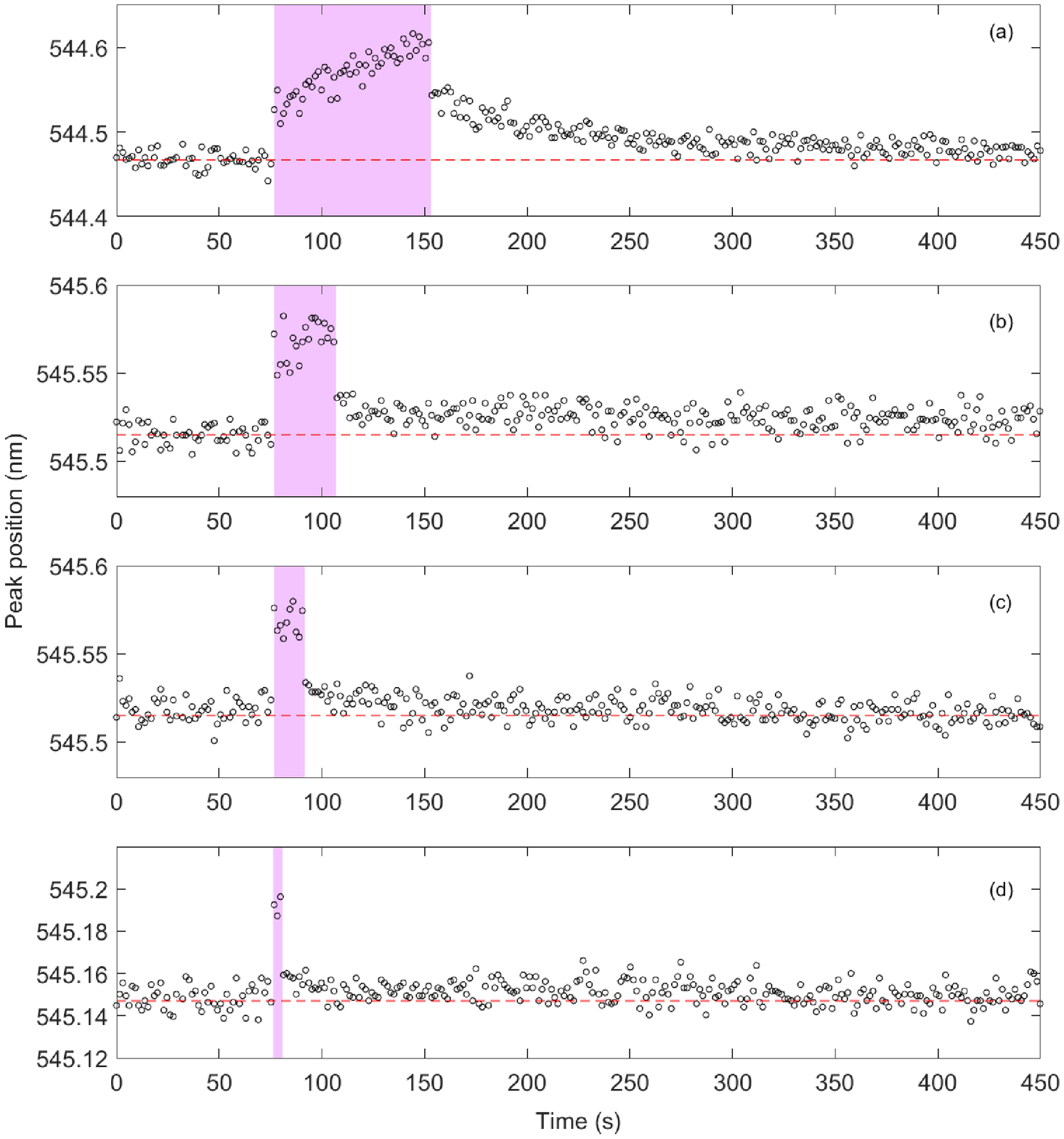} 
 	\end{center}
 
 		 \caption{(Color online) Time-traces of the center wavelength of QDs PL peak during 76.5s (a), 30 s (b), 15 s (c), and 4.5 s (d) plasma exposure at input power of 50 W and pressure of 4 Pa. The shaded region indicates the plasma exposure.} \label{f6}
 \end{figure}
 
\subsection{Slow shift}
\subsubsection{Thermal effects}
 First, the long timescale drift of the center wavelength of the PL peak is mainly attributed to thermal effects of the plasma on the substrate and the QDs. It is expected that the temperature-induced enhancement of the lattice dilation and the interaction between excitons and phonons lead to the PL redshift \cite{varshni1967temperature}.

  We also measured the temperature of the QD sample during this plasma exposure (see Fig.3-b) and observed, on average, a redshift of 0.11 nm for an increase of 1.1 K in sample temperature during plasma exposure. This redshift in the PL wavelength of CdSe/ZnS QDs due to increasing temperatures around ambient conditions has been well established in literature and comparable to \cite{valerini2005temperature,walker2003quantum}. 
  We cross-checked this effect by externally heating the QDs sample without applying plasma and recorded comparable redshifts (about 0.1 nm/K) of the center wavelength of the PL peak as a function of sample temperature (See Fig.1 in SM).

  Next to the wavelength shift, we also observed a decrease in the PL intensity during plasma exposure (red circles in Fig3-a). This is also mostly attributed to thermal effects. From literature it is well known that thermal activation of surface trapped states and increased non-radiative recombinations may lead to a decrease in PL intensity \cite{valerini2005temperature}. 
  We quantify this thermal effect on the PL intensity with the external heating to be $- 1.6~\%/$K (See Fig.1 in SM). This is in agreement with the literature values reporting a temperature-related PL intensity reduction of $- 1.6~\%/$K which was obtained for the heating of CdSe/ZnS QDs in poly(lauryl methacrylate) matrices \cite{walker2003quantum}.
  
\subsubsection{Ion effects} 
The decrease in the PL intensity during plasma exposure is slightly higher, i.e. $-1.9~\%/$ K. This difference is most likely due to ion-induced effects on the PL emission of the QDs, as we will explain below. The intensity largely recovers when the QDs cool down after plasma exposure, but not fully due to this additional ion-induced effect.
    
The plasma-enabled ion bombardment might partly sputter the ZnS shell and therefore the PL emission intensity may be reduced, being essentially a non-reversible effect. In theory, this could be a reason for the fact that the PL emission intensity reduction is not fully reversible in the case of 76.5 s plasma exposure. However, this effect cannot change the exciton localization and the PL wavelength; if the ion sputtering could impact until the CdSe core, quantum confinement would be lost and the specific QD would be expected to no longer contribute to the overall PL signal.
   
 Another plasma-induced effect that can influence the PL properties of QDs is the incorporation of plasma ions in the QDs. Due to ion incorporation into the semiconductor lattice, point defects may possibly be introduced, consequently affecting the optical property of the semiconductor \cite{schoenfeld1998argon}. These ion-induced defects can for instance serve as local non-radiative recombination centers for electron–hole pairs, reducing radiative lifetimes and the PL efficiency and intensity.

   The above mentioned effects may contribute to the "slow shift", but their time scales (see section 
   B in SM) are too long to explain the initial "fast shift". Figure \ref{f6} shows the time-resolved PL peak position of the QD samples for different plasma exposure times while all other plasma parameters were kept the same for these measurements. As it can be seen from figure \ref{f6}, by decreasing the plasma exposure time from 76.5 s to 4.5 s, while the initial fast shift remains the same at 0.04 nm, the temperature-induced  shift reduces and become negligible for very short plasma exposures of 4.5 s (see fig. \ref{f6}-d). Moreover, only for long plasma exposures, there remains a small (about $1\%$) non-recovered portion in the PL peak position 450 s after switching off the the plasma (see fig. \ref{f6}-a and fig. \ref{f6}-b). For 4.5 s of plasma exposure, there remains only a fast redshift that completely recovers (see fig. \ref{f6}-d). This shows that the ion-induced effects do not have any influence on the PL emission of QDs at very short plasma exposure.
    
  \subsection{Fast shift; electric field and charge effects} 
 This "fast shift" appears to be an immediate result of direct plasma impact. Here, the QDs will undergo a redshift ($\Delta{\lambda}$) under the influence of the charges and electric fields locally near their surface due to the quantum-confined Stark effect \cite{ekinov1990quantum, seufert2001stark}. This redshift is expressed as \cite{efros2018evaluating}:
 
  \begin{eqnarray}
&&{\Delta{\lambda}}=0.03{\lambda^2}({hc})^{-1}\left({m_e}^*+{m_h}^*\right){a^4}\left(\frac{2 \pi eE}{h}\right)^2,
\end{eqnarray}

where $e$, $h$, $a$, ${m_e}^*$ and ${m_h}^*$ indicate the elementary charge, Planck's constant, the QD's core radius, the electron's effective mass, and the hole's effective mass, respectively. 
In our situation, the observed Stark shift can be either due to charge carriers present on the surface of the QDs or on the substrate surface near QDs \cite{wang2001calculating,early2010polarization, yalcin2011spectral}, or due to plasma  self-induced macroscopic electric fields present in the vicinity of the QDs \cite{seufert2001stark}.

\begin{figure}
 	\begin{center}
 		\includegraphics[width=1\textwidth]{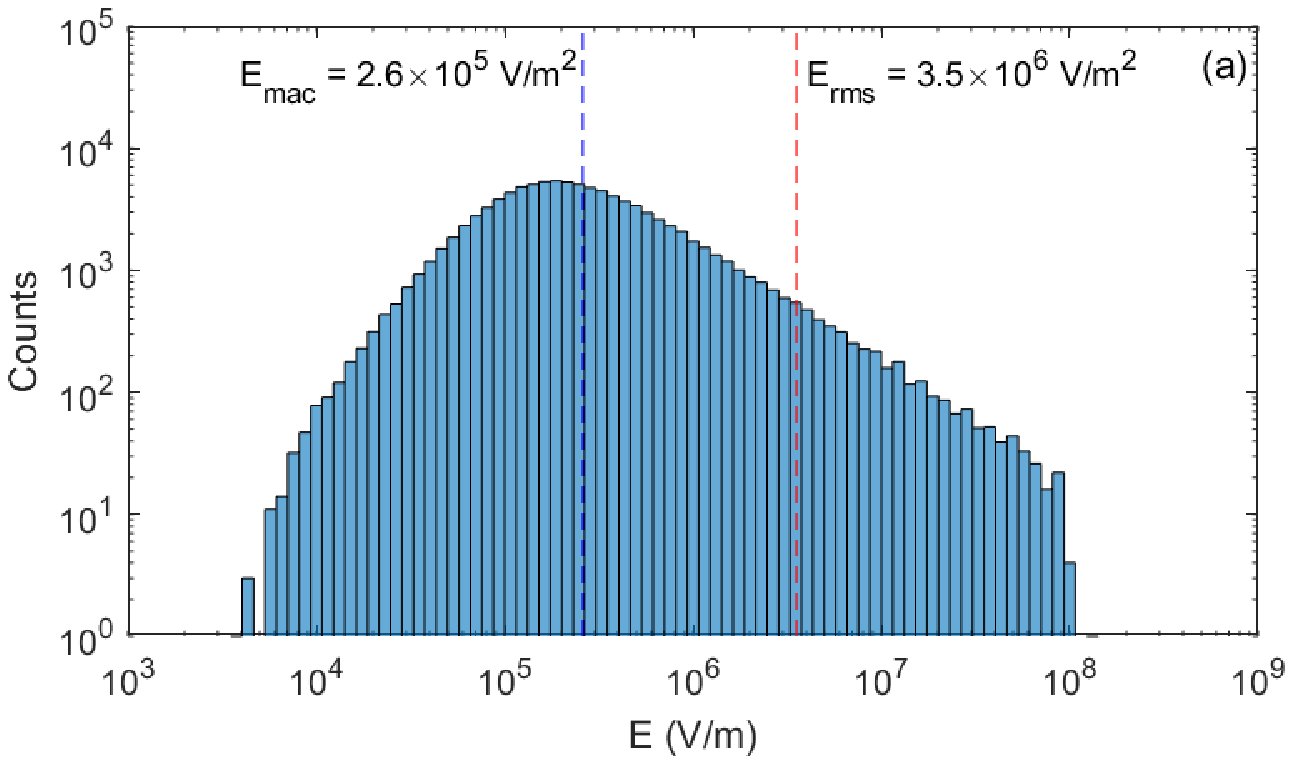} 
	\end{center}
 	\begin{center}
 		\includegraphics[width=1\textwidth]{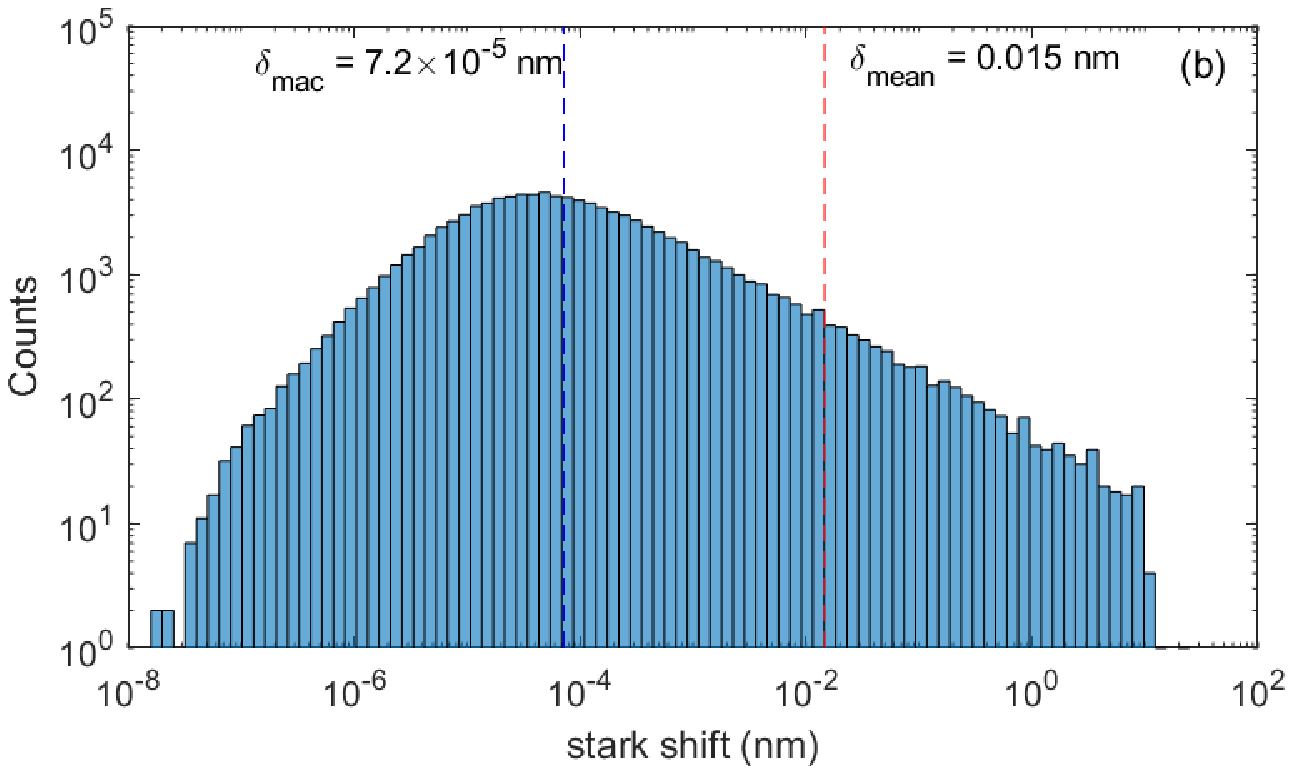} 
 	\end{center}
 \caption{(Color online) Histogram of the microscopic electric field E$_{mic}$ at the location
of a QD on the substrate surface (a) and the relevant histogram of QD photoluminescence Stark shifts integrated over $10^{5}$
different quasi-stationary electron configurations.} \label{f7}
 \end{figure}

First we consider the possible quantitative influence of the macroscopic electric field ($E_{mac}$) - i.e. the plasma sheath electric fields - on the PL peak's center wavelength of the QD.
The estimated plasma-induced macroscopic electric field is approximately 2.6$\times10^{5}$ V/m at an electrically floating surface near an RF sheath edge under the current plasma conditions \cite {barnat2007electric, kim2003simulation}. This would lead to a Stark redshift of $\delta_{mac} = \Delta\lambda (E_{mac})$ = $7.2\times10^{-5}$ nm (Eq.1), much smaller than the observed fast redshift of 0.04 nm. Therefore, the macroscopic electric field near the sample's surface cannot directly explain the observed spectral fast shift.


The macroscopic electric field induces a surface charge on the substrate. This surface charge randomly fluctuates due to the discrete nature of charging process \cite{sheridan2011charge,heijmans2016dust}. It can locally cause a larger microscopic electric field which can induce a larger Stark shift. To evaluate this effect, we consider a spherical QD located on a Si substrate sensing a microscopic electric field caused by charges present on the substrate surface facing the plasma.
The Stark shift due to the microscopic electric field (near the QD) induced by a certain quasi-static electron configuration on the surface is calculated using Eq.1. Since, the experimentally observed ${\Delta{\lambda}}$ has been measured with 750 ms exposure time, it is integrated over many quasi-static electron configurations as well as many photoluminescence liftimes of the QDs. To account for this, we first calculate the spectral shift for each quasi-static electron configuration and then average the shift over all configurations.

The surface charge density due to the macroscopic electric field of $2.6\times10^{5} $ V/m  distributed over the substrate surface as $10^{5}$ random quasi-static electron configurations. The microscopic electric fields and their fluctuations calculated at the location of the QD ($a$ = 2.1 nm) for each quasi-static electron configuration. The histogram of the microscopic electric field and the relevant stark shift are shown in figure \ref{f7}. The averaged shift is indicated as mean of the histogram, $\delta_{mean} = \Delta\lambda (E_{rms})$ = 0.015 nm (see Fig.5-b).
 This shift is close to our observed fast redshift (0.04 nm). The difference may be explained by the influence of charges directly attached to the QD's surface which is not involved in the calculation of the Stark shift in the above surface charge model. To account for this, we will evaluate the contribution of direct charging of the QDs by the plasma as follows.     


The charge on a spherical particle on a flat surface facing a plasma imposing an electric field $E$, can be obtained from $Q = (1.64)(4\pi\epsilon_0)Ea^{2}$, with $a$ being the particle radius, known as the "shared charge model" \cite{wang2007charge}. Considering the macroscopic electric field of 2.6$\times10^{5}$ V/m, the calculated shared charge will be about 0.002 e. 
This means that on average $ 0.2 \%$ of QDs on the sample are charged by one elementary charge. 
According to the results of the quantum-confined Stark shift models \cite{early2010polarization, seufert2003single}, an electron on the QD's surface - of the size and material as used in this work - would induce a redshift of 14 nm of the PL peak position. Considering the measured PL spectrum as the sum of both unshifted PL spectra emitted by uncharged QDs (most of the QDs in the population) and 14 nm redshifted spectra emitted by charged QDs ($ 0.2 \%$ of the total QD population as explained above), the resulted stark shift will be 0.02 nm which is also in a good agreement with our observed fast redshift. However, the assumption of charged and uncharged QDs on the surface is over-simplified. In reality, for instance, uncharged QDs in the neighborhood of charged QDs will also be influenced by Stark effects induced on longer length scales.

From the analyses above, we can conclude that the observed fast redshift can be explained by the total Stark shift including both contributions, the 0.015 nm redshift due to the charges near the QDs on the substrate surface and the 0.02 nm redshift due to the charges attached to QDs under the current plasma conditions. With our technique we were able to experimentally determine a Stark redshift of 0.04 nm of the PL peak center wavelength of the QDs. From such measurement, one could estimate for instant (plasma-induced) surface charges and the charge on nanoparticles on plasma-facing surfaces. 

\section{conclusion}
In conclusion, we demonstrated for the first time that it is
possible to directly detect plasma charging of nanometer-sized structures by measuring the time-dependent, charge-induced, spectrally-shifted photoluminescence signal emitted by QDs on an electrically floating substrate facing a plasma environment. We measured a total maximum spectral red shift of 0.15 nm with an accuracy of 0.02 nm, and were able to differentiate between a direct plasma charging induced fast shift and a thermal- (plasma-induced) and ion-induced shift on longer time scales. Ultimately, the method introduced here may enable charge measurements of airborne QDs inside the plasma bulk volume, which could even be calibrated absolutely by applying laser-induced photodetachment \cite{10.1088/1361-6463/ac1761} and additional sensitive plasma diagnostics such as microwave cavity resonance spectroscopy (MCRS) \cite{beckers2018mapping, platier2019resonant}. The current work and the method it introduces allows the research community to shine new light on the most fundamental processes regarding the charge collection of micro and nanostructures in various fields, ranging from complex and dusty plasma physics to aerosol science and astrophysics. 

\section*{ACKNOWLEDGEMENT}
This research was supported by the Dutch Research Council, NWO (project number 15710). The authors are particularly grateful to Dr. Mikhail Pustylnik and Prof. Sangam Chatterjee for the valuable and fruitful discussions.    



\section*{REFERENCES}

\bibliography{arXivPreprintZahraMarvi}

\clearpage

\end{document}